\documentclass[twocolumn]{aastex631} 

\usepackage{hyperref}

\defcitealias{rosas-guevara_evolution_2022}{RG22}

\shorttitle{AASTeX v6.3.1 Sample article}
\shortauthors{Ash et al.}
\graphicspath{{./}{figures/}}

\begin{document}

\title{Stellar bars form dark matter counterparts in TNG50}

\author[0009-0003-7613-3109]{Neil Ash}
\affiliation{University of Michigan Department of Astronomy \\
1085 S. University \\
Ann Arbor, MI 48109, USA}

\author[0000-0002-6257-2341]{Monica Valluri}
\affiliation{University of Michigan Department of Astronomy \\
1085 S. University \\
Ann Arbor, MI 48109, USA}

\author[0000-0002-5970-2563]{Yingtian Chen}
\affiliation{University of Michigan Department of Astronomy \\
1085 S. University \\
Ann Arbor, MI 48109, USA}

\author[0000-0002-5564-9873]{Eric F. Bell}
\affiliation{University of Michigan Department of Astronomy \\
1085 S. University \\
Ann Arbor, MI 48109, USA}

\begin{abstract}

Dark matter (DM) bars that shadow stellar bars have been previously shown to form in idealized simulations of isolated disk galaxies. Here, we show that DM bars commonly occur in barred disk galaxies in the TNG50 cosmological simulation suite, but do not appear in unbarred disk galaxies. Consistent with isolated simulations, DM bars are typically shorter than their stellar counterparts and are $75\%$ weaker as measured by the Fourier $A_2$ moment. DM bars dominate the shape of the inner halo potential and are easily identified in the time series of quadrupolar coefficients. We present two novel methods for measuring the bar pattern speed using these coefficients, and use them to make a measurement of the pattern speed and rotation axis orientation for one sample galaxy located in one of the high-time resolution subboxes of TNG50. The stellar and dark bars in this galaxy remain co-aligned throughout the last 8 Gyr and have identical pattern speeds throughout. Both the pattern speed and rotation axis orientation of the bars evolve considerably over the last 8 Gyr, consistent with torques on the bars due to dynamical friction and gas accretion. While the bar pattern speed generally decreases over the time course, it is seen to increase after mergers. Rather than remaining static in time, the rotation axis displays both precession and nutation possibly caused by torques outside the plane of rotation. We find that the shape of the stellar and DM mass distributions are tightly correlated with the bar pattern speed.
\end{abstract}


\section{Introduction} \label{sec:intro}

It has been well established that observed stellar bars are close to maximally rotating ($r_{bar}/r_{corot} \equiv \mathcal{R} \approx 1$) \citep[e.g.][]{corsini_direct_2010,aguerri_bar_2015,cuomo_bar_2019,guo_sdss-iv_2019} while simulated bars are not, presenting a challenge to $\Lambda$CDM \citep[e.g.][]{algorry_barred_2017, roshan_fast_2021, ansar_bar_2023}. One contributor to this tension is the observed slowdown of simulated bars caused by the transfer of angular momentum to the dark matter (DM) halo through dynamical friction with material in resonance with the bar \citep[e.g.][]{tremaine_dynamical_1984,debattista_constraints_2000,athanassoula_bars_2005}. These interactions cause the bar to both grow in strength and slow over time. Stellar bars in isolated simulations have also been found to trap DM into bar-like orbits, forming a DM ``halo'', ``shadow'', or ``ghost'' bar \citep{athanassoula_bars_2005,berentzen_growing_2006,colin_bars_2006, athanassoula_bar_2007, athanassoula_bar_2013, saha_spinning_2013, petersen_dark_2016, collier_coupling_2021, marostica_response_2024}. \cite{colin_bars_2006}, \cite{berentzen_growing_2006}, and \cite{athanassoula_bar_2007} claim that these DM bars lag behind their stellar bar by $\lesssim 10$ degrees, creating a torque on the stellar bar which enhances the rate of pattern speed slowdown. 
In contrast, \cite{petersen_dark_2016} find that the trapping of DM in bar-like orbits reduces the rate of angular momentum transport to the outer halo from the stellar bar, thereby reducing the rate at which the bar pattern speed decreases. 

Because of the apparent tension between observed and simulated bars and the importance of bar-halo interactions on the growth and evolution of stellar bars, it is both valuable and timely to pursue a detailed understanding of the response of the inner halo to the stellar bar within fully cosmological $\Lambda$CDM simulations, where hierarchical galaxy assembly, gas physics, satellite mergers, and baryonic feedback are each modelled.
In this paper, we demonstrate that DM bars are common in the TNG50 cosmological simulation, matching predictions from isolated, idealized $N$-body disk simulations. The presence of these halo bars becomes evident in the time series quadrupolar Basis Function Expansion (BFE) representations of either the DM density or potential. BFE models provide a rich representation of the bar structures and their evolution, and enable a measurement of both the pattern speed and rotation axis evolution provided sufficiently fine time resolution $\lesssim 10$ Myr.

\section{Simulations}\label{methods:simulations}

The simulation used in this work is TNG50 of the IllustrisTNG simulation suite \citep{pillepich_first_2019,nelson_first_2019,nelson_illustristng_2019}. TNG50 is a magnetohydrodynamical $N-$body simulation run using the moving mesh $\texttt{AREPO}$ code \citep{weinberger_arepo_2020}. It evolves a box of side length $\sim 50$ cMpc from $z=127$ to $z=0$, with a DM mass resolution of $\sim 3.1\times10^5 h^{-1} M_\odot$ and mean baryonic mass resolution of $5.7\times10^4 h^{-1} M_\odot$. The simulation adopts the Planck 2015 cosmology with $h=0.6774$ \citep{planck_collaboration_planck_2016}. We make use of Subbox-0, a sub-volume within TNG50 of side length $\sim 7.5 h^{-1}$ cMpc, and the suplementary data catalog introduced in \cite{nelson_first_2019}. Subbox-0 is a relatively dense environment containing $\sim 6$ Milky Way-mass halos. It provides $\sim 3600$ snapshots, offering a time resolution of $\sim 8$ Myr near $z=0$ allowing bar rotation to be fully resolved. Finally, we make use of the catalogue of TNG50 barred galaxies and their properties produced by \cite{rosas-guevara_evolution_2022} (hereafter \citetalias{rosas-guevara_evolution_2022}).

\section{Methods} \label{sec:methods}

We use several methods in this work which have been well established. Determination of principal axis lengths $a,b,c$ and orientation is performed using the iterative shape-tensor method \citep[e.g.][give our exact conditions to terminate iteration]{emami_morphological_2021, ash_figure_2023}.

To assess whether DM halo bars are common in TNG50 barred galaxies, we take the catalog of $349$ galaxies at $z=0$ presented in \citetalias{rosas-guevara_evolution_2022}. This catalog contains massive disk galaxies with stellar masses $M^* > 10^{10} M_\odot$, of which $\sim 30\%$ are barred. Using this galaxy set, we compute the in-plane Fourier $m=2$ amplitude $A_2(r)$ at $z=0$ for both the stellar and DM particle distribution, using the definitions defined by \citetalias{rosas-guevara_evolution_2022} and cylindrical bins of width 0.1 kpc and height 1 kpc centered on the mid-plane and potential minimum. 

The presence of a stellar bar leaves an aspherical imprint on the potential and density which is well encoded by the spherical harmonic BFE models. We compute these models using the \texttt{AGAMA} software package \citep{vasiliev_agama_2019}. Our BFE models use $N_{grid}=25$ logarithmically-spaced concentric shells centered on the potential minimum and $l_{max}=m_{max}=6$ for our limiting expansion order. We perform this expansion in the inertial simulation box frame without any coordinate rotation. 

We use two novel methods to measure bar pattern speed using the time series of the quadrupolar BFE coefficients. These methods are described in full elsewhere \citep[\textit{in prep.}]{ash_basis_2024}. We provide a brief introduction to each of these below.

\subsection{Continuous Wavelet Transform}\label{methods:CWT}

The rotation of the bar leaves an oscillatory signal in the quadrupolar BFE coefficients which is related to the pattern speed. To determine the time dependence of frequencies represented in our BFE coefficients, we utilize a Continuous Wavelet Transform (CWT). A CWT is the convolution of an input signal $x(t)$ with a wave function $\psi(t)$: \begin{equation}
    X(a,b) = \frac{1}{|a|^{1/2}} \int_{-\infty}^\infty x(t) \bar{\psi}\left(\frac{t-b}{a}\right),
    \label{CWT}
\end{equation}
where $X(a,b)$ is the CWT of the input signal, $a$ specifies a ``scale'' value, and $b$ specifies the center for the convolution. In contrast to a Fourier transform, in CWT $\psi(t)$ represents a wavelet for which $\lim_{|t|\to \infty}\psi(t) = 0$. We choose a complex Morlet wavelet \citep{grossmann_decomposition_1984}, which defines a wavelet using a complex exponential with a Gaussian window function:
\begin{equation}    \Psi(t) = \frac{1}{\sqrt{\pi B}}\exp\left(\frac{-t^2}{B}\right)\exp(i2\pi C t).
    \label{Morlet_wavelet}
\end{equation}
The value $B$ representing the width of the Gaussian is equivalent to twice the variance, and the value $C$ gives the central frequency of the complex exponential. We find that choosing $B=10$ and $C=1$ gives a good trade-off between precision in the time and frequency domains and adopt these values for this work.


CWT requires uniform time sampling. We achieve this by taking BFE quadrupolar coefficients generated from the subbox snapshots (which do not have a uniform time spacing) and linearly interpolating between snapshots on a grid with spacing $\Delta t \approx 4.8$ Myr. We use a grid of 200 scale values logarithmically spaced between the Nyquist frequency ($a=2$, $f\sim105$ Gyr$^{-1}$) and the frequency corresponding to one full period in our full 8 Gyr timecourse ($a=1679$, $f\sim 1/8 $ Gyr$^{-1}$). We note that, since the value of the quadrupolar spherical harmonics goes through two full periods for one complete rotation in either azimuth or polar angle, the physical frequency of a rotating object is half the frequency observed in its quadrupole coefficients. We therefore divide the frequencies obtained in the CWT analysis by 2. We make use of the $N$-dimensional CWT implemented in the $\texttt{pywt}$ software package \citep{lee_pywavelets_2019} to simultaneously perform a CWT over all radii and quadrupolar orders $l=2, -2\leq m \leq 2$.



\subsection{Rotation axis and pattern speed fitting}\label{methods:WignerDmat}

The rotation of a spherical harmonic function $Y_m^l$ can be expressed as a linear combination of up to $2l+1$ other spherical harmonic functions of the same order $l$. The weighting coefficients to perform the rotation are taken from the Wigner $\mathcal{D}$-matrices:
\begin{equation}
    Y_m^l(\mathbf{R}^{-1}\mathbf{n}) = \sum_{m'=-l}^{l} \mathfrak{D}^{(l)}_{mm'}[\mathbf{R}]Y^l_{m'}(\mathbf{n}),
    \label{eq:WignerD_rotation}
\end{equation}
Where $\mathbf{n}$ is a unit vector within the non-rotated basis and $\mathbf{R}$ is either a rotation matrix or unit quaternion describing the rotation. The matrix elements $\mathfrak{D}^{(l)}_{mm'}[\mathbf{R}]$ are defined as:
\begin{equation}
    \mathfrak{D}^{(l)}_{mm'}[\mathbf{R}] \equiv \left<lm|\mathbf{R}|lm'\right>.
    \label{eq:D-matrix_elements}
\end{equation}

We use this property to define a method to fit simultaneously the axis of rotation and pattern speed for the stellar bar directly from its quadrupolar BFE coefficients, described in full by \citep[][\textit{in prep.}]{ash_basis_2024}. In brief, this method attempts to determine the co-rotating frame in which the time variation of the quadrupolar $l=2$ coefficients is minimized. We measure this time variation as:
\begin{equation}
  \chi^2 = \frac{1}{W}\sum_{i,j,m} \frac{w_j\left\{\rho_l^m(t_{j-1},r_i) - \mathbf{R}\rho_l^m(t_j,r_i) \right\}^2}{\sigma^2[\rho_l^m(t_j,r_i)]}\bigg|_{l=2},
    \label{eq:WignerD_chi2}
\end{equation}
where $w_j$ are weighting coefficients for which $\sum_j w_j = W$, $t_j$ and $t_{j-1}$ are subsequent simulation snapshots, $r_i$ is a radial grid node, $\sigma^2[\rho_l^m(t_j,r_i)]$ is an estimate of the squared coefficient uncertainty, and
\begin{equation}
    \mathbf{R}\rho_l^m(t_j,r_i) = \sum_{m'=-l}^l\rho_l^{m'}(t_j,r_i)\mathcal{D}^{(l)}_{mm'}[\mathbf{R}(t_j)]
    \label{rot_coeffs}
\end{equation}
represents the new BFE coefficient value $\rho_l^m$ at snapshot $t_j$ after the rotation $\mathbf{R}(t_j)$ is applied. $\mathbf{R}(t_j)$ is a function of three parameters:
\begin{equation}
    \mathbf{R}(t_j) = \mathbf{R}[\theta,\phi,\alpha(t_j)],
\end{equation}
Where $\theta$ and $\phi$ specify the polar and azimuthal angle of the axis of rotation (respectively, in the inertial simulation box frame) and $\alpha(t_j)$ gives the angle of rotation, which is expressed in terms of a fixed pattern speed as 
\begin{equation}
    \alpha(t_j) = \Omega_p \cdot (t_j - t_{j-1}).
\end{equation}
The summation in equation \ref{eq:WignerD_chi2} may be performed with an arbitrary range of snapshots and radial grid nodes. In the limit that a majority of the time evolution of the BFE coefficients is explained by a 3D rotation (as is the case for a stellar bar rotating as a solid body), our function $\chi^2$ is minimized when $\mathbf{R}(t_j)$ matches the system rotation and each of the coefficients are rotated into the co-rotating frame. Determining the rotation of a system is therefore reduced to a minimization problem of three variables; $\theta$ and $\phi$ describing the orientation of the rotation axis, and $\Omega_p$ describing the pattern speed/ rotational velocity, and can be performed using standard $\chi^2$-minimization techniques.


To allow for a time-varying pattern speed, we perform a fit for the rotation properties at each snapshot with a window filter 30 snapshots across and centered on the current snapshot. We further adopt Gaussian weights $w_j$ centered on the current snapshot and with a standard deviation of $\sim 42$ snapshots ($\sim 290$ Myr for the median snapshot spacing, $\sim 360$ Myr near $z=0$ and $\sim 200$ Myr near $z=2$). We choose these weights to approximately correspond to the Gaussian kernel used for our CWT analysis, near the typical bar pattern speed of $30 - 60$ km s$^{-1}$ kpc$^{-1}$. We restrict the radial extent of our fitting to the bar region, from $ 0.2 \lesssim r \lesssim 1 $ comoving $h^{-1}$ kpc. By performing our fitting routine in this manner, we are able to take advantage of the extra information provided by many snapshots' BFE coefficients while also approximating the time-variation of the rotation axis and pattern speed. 

Our fit results do not depend on whether we use the coefficients from the expansion of the density or the potential. The width of the Gaussian kernel and window function additionally do not affect the mean value of our fits but instead act to suppress the influence of noisy snapshots. Finally, the exact value we assume for $\sigma^2[\rho_l^m(t_j,r_i)]$ does not impact our fit results (since we assume it to be constant), but rather helps the minimization routine to converge. We assume $\sigma[\rho_l^m(t_j,r_i)]$ to be $10\%$ of the 99$^\mathit{th}$ percentile of the absolute value of the quadrupolar coefficient values. For a more in-depth investigation of the BFE coefficient uncertainties, see \cite[][, in prep.]{ash_basis_2024}. To calculate the rotated coefficient modes, we make use of the python software packages \href{https://quaternionic.readthedocs.io/en/latest/}{\texttt{quaternionic}} and \href{https://pypi.org/project/spherical/}{\texttt{spherical}} developed by Michael Boyle.

\section{Results} \label{sec:results}

\subsection{Prevalence of DM bars}
\begin{figure*}
    \centering
    \includegraphics[width=.85\textwidth]{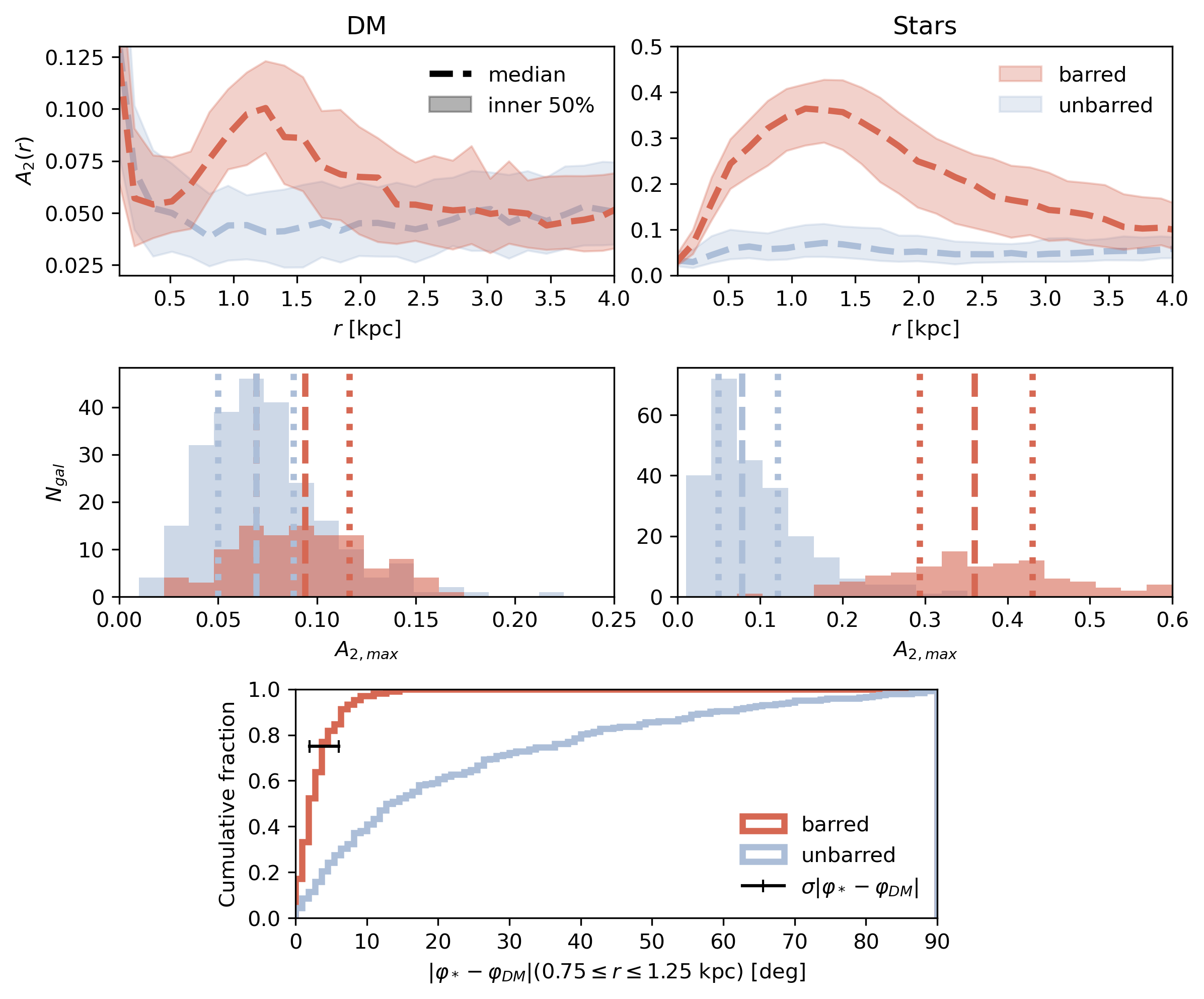}
    \caption{Bar metrics applied to the \citetalias{rosas-guevara_evolution_2022} disk galaxy catalog for DM (left) and stellar (right) particle distributions. Top row: $A_2(r)$, with the median (dashed line) and interquartile ranges (shaded region) shown for both the barred (red) and unbarred (blue) populations as defined by \citetalias{rosas-guevara_evolution_2022}. The presence of a stellar bar is easily identified by the peak in $A_2(r)$. In the DM, the median $A_2(r)$ curves show that the DM in barred galaxies is typically elongated within $\sim 2 $ kpc more strongly than in unbarred galaxies. Middle row: histograms for $A_{2,max}$. Medians (dashed lines) and the interquartile region (dotted lines) of the distributions are shown. The ratio of stellar bar to DM bar strength is approximately $4:1$ for the median barred galaxy in this catalog, as measured by $A_{2,max}$. Bottom row: cumulative histogram of the difference in phase angles $\varphi(r)$ between the stellar and DM mass distributions, measured where the bar amplitudes are typically highest. Typical measurement uncertainty is indicated by the black line. The distribution of $|\varphi_{*}-\varphi_{DM}|$ is consistent with no net offset between stellar bars and their DM counterparts.
    }
    \label{fig:RG22_A2_stats}
\end{figure*}

 In figure \ref{fig:RG22_A2_stats}, we show the median and interquartile region for the amplitude of $A_2(r)$ for both stars and DM in the barred and unbarred populations. The presence of a bar is clear in the stellar population, with a median peak amplitude (defined here as the amplitude at the first maximum determined by a cubic spline outside 0.3 kpc) of $A_{2,max}\sim 0.36$, as compared to the median peak in the unbarred population of $A_{2,max}\sim 0.08$. While the DM of the barred galaxies features a lower $A_{2,max}$ of $\sim 0.09$, there is a systematic offset in the DM $A_2(r)$ between the barred and unbarred galaxies. By taking a ratio of $A_{2,max}$ for DM and stellar bars, we find that the ratio of DM to stellar bar strength is $\sim 0.26$ for the median barred galaxy in our sample.


The elongation in the DM distribution is typically aligned with the major axis of the stellar bar, as revealed by the phase angles of the DM and stellar $m=2$ moment $\varphi(r)$. We measure the difference in phase angles $|\varphi_* - \varphi_{DM}|$ using all particles with $|z|\leq 0.5$ and within the the radial range $0.75 \leq r \leq 1.25$ kpc, where both DM and stellar bar strengths peak for the median barred galaxy. We find no evidence for a net offset between the DM and stellar bars at a population level: In barred galaxies, $|\varphi_* - \varphi_{DM}|$ is less than our typical measurement uncertainty ($\sim 4^\circ$, estimated by 100 bootstrap iterations) in $71\%$ of barred galaxies and less than twice our uncertainty in $94\%$ of barred galaxies. This alignment of the $m=2$ Fourier modes is only weakly observed in unbarred galaxies, which instead show a nearly uniform distribution in phase angle. We note that the maximum possible separation of phase angles is $90^\circ$, because the $m=2$ mode is symmetric for reflections about the origin. 

Taken together, we can observe that the DM distribution in the inner regions of barred galaxies is systematically more elongated than in unbarred galaxies, and that this elongation is aligned to the stellar bar major axis. The presence of DM bars in the TNG50 cosmological $\Lambda$CDM simulation is consistent with previous results from a large body of idealized $N$-body simulations of isolated galaxies \citep{athanassoula_bars_2005,berentzen_growing_2006,colin_bars_2006,athanassoula_bar_2007,saha_spinning_2013,petersen_dark_2016,collier_coupling_2021,marostica_response_2024}, however the degree of alignment between the DM and stellar bars is inconsistent with \cite{athanassoula_bar_2007} who finds the two bars to be misaligned by $\lesssim 10^\circ$ in her simulations.


\subsection{Detailed investigation of SubhaloID 574286 in Subbox-0 }

\begin{figure*}
    \centering
    \includegraphics[width=\textwidth]{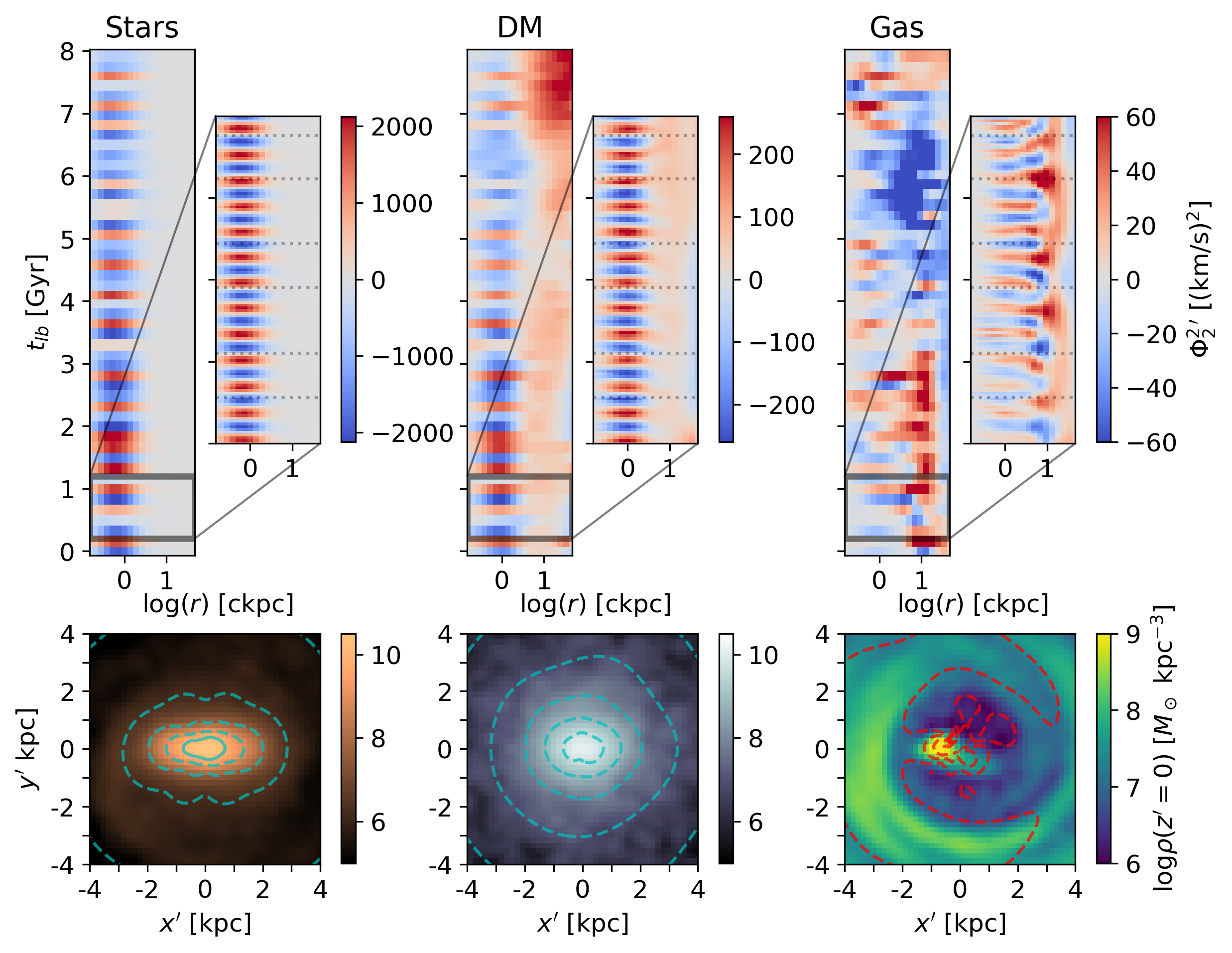}
    \caption{Contributions to the potential from stellar (left columns), DM (center columns), and gas (right columns) particles for TNG50 subhaloID 574286. Top row: rotated $l,m=2$ quadrupolar coefficients for the 'coarse' time resolution main TNG50 box (left hand side of each column) next to the high time resolution Subbox-0 snapshots between $0.2 \leq t_{lb} \leq 1.2$ Gyr (right hand side of each column, tick intervals are 250 Myr) for the same halo. The location of the 'coarse' time resolution snapshots are shown here as dashed horizontal lines. 
    Aliased oscillations caused by the bar's rotation are visible in the 'coarse' time resolution snapshots, but can only be resolved using the Subbox-0 snapshots. Bottom row: contours for the BFE density model of each particle type in the region near the bar ($r \lesssim 4$ kpc) are shown overplotted on the mid-plane mass density at $z\sim 0$. Each of the plots are taken in primed coordinates rotated such that the bar major axis lies along $x'$, and minor axis along $z'$ (out of the page). The DM density is aspherical and elongated in the direction of the stellar bar. 
    }
    \label{fig:l2_coeffs_and_contours}
\end{figure*}

The ratio of $A_2(r)$ Fourier amplitude between barred and unbarred DM distributions is considerably smaller relative to the same ratio for stellar distributions. Attempting to detect the presence of a DM bar using $A_2(r)$ is therefore challenging. In contrast, we find DM bars are easily detectable in the time series of the quadrupolar BFE coefficients. In these coefficients, the DM bar produces a semi-sinusoidal oscillation in the inner regions as the bar structure comes into and out of phase with a given spherical harmonic basis function. We chose one galaxy (SubhaloID 574286), which belongs to the \citetalias{rosas-guevara_evolution_2022} catalogue and lies within Subbox-0 for the last $\sim 8$ Gyr, for closer study. In figure \ref{fig:l2_coeffs_and_contours}, we show the $l,m=2$ coefficient values across time and radius for the gas, star, DM contributions to the potential of this galaxy. The coefficients are rotated to approximately align the rotation axis to the $z$-axis, which places the azimuthal oscillations of the $m=2$ spherical harmonic within the plane of rotation.

The presence of the bar is apparent in the inner regions of the galaxy as unresolved oscillations in the coefficient amplitudes. These oscillations are fully resolved in the Subbox-0 snapshots (shown in the inset plots in fig \ref{fig:l2_coeffs_and_contours}). Notably, the coefficients of the stellar and DM potentials share a common frequency and are in phase with one another. The gas coefficients do not follow this frequency, but instead appear to periodically extend into the bar region from larger radii. This could be indicative of gas accretion into the bar region. In the bottom row of figure \ref{fig:l2_coeffs_and_contours}, we show contour plots of the 3D density contributions in the $x-y$ plane from each of the three particle types at present day, rotated to align the bar major axis to $x$ and rotation axis to $z$. Density contours from our BFE models are overplotted. The DM density is elongated in the same direction as the stellar bar. This is reflected by the peak amplitudes of the $l,m=2$ potential coefficients, which act as a correction to a spherical potential and are a factor $\sim 9$ lower in the DM compared to the stellar coefficients. This is consistent with the lower $A_{2,max}$ amplitudes for DM bars compared to stellar bars shown in figure \ref{fig:RG22_A2_stats}, nonetheless the presence of the DM bar is easily identified by the oscillation in the quadrupolar coefficients. The contour plots of the gas density show a trailing-arm like feature beginning at the outer edge of the bar and extending for several kpc. 

We see evidence for DM ``wake'' structures in the BFE coefficients. These structures emerge just beyond the bar region as faint filaments which extend to larger radii and are sloped generally downwards. This downward slope implies that they reach a given phase angle after the DM/stellar bar, and that the time at which they arrive at this phase angle is radius dependent. This picture is consistent with that of e.g. \cite{athanassoula_bars_2005} and \cite{petersen_dark_2016}, who show that DM wakes may trail behind stellar bars and exert a negative torque on them. 

\subsubsection{Pattern speed and rotation axis evolution}

\begin{figure*}
    \centering
    \includegraphics[width=\textwidth]{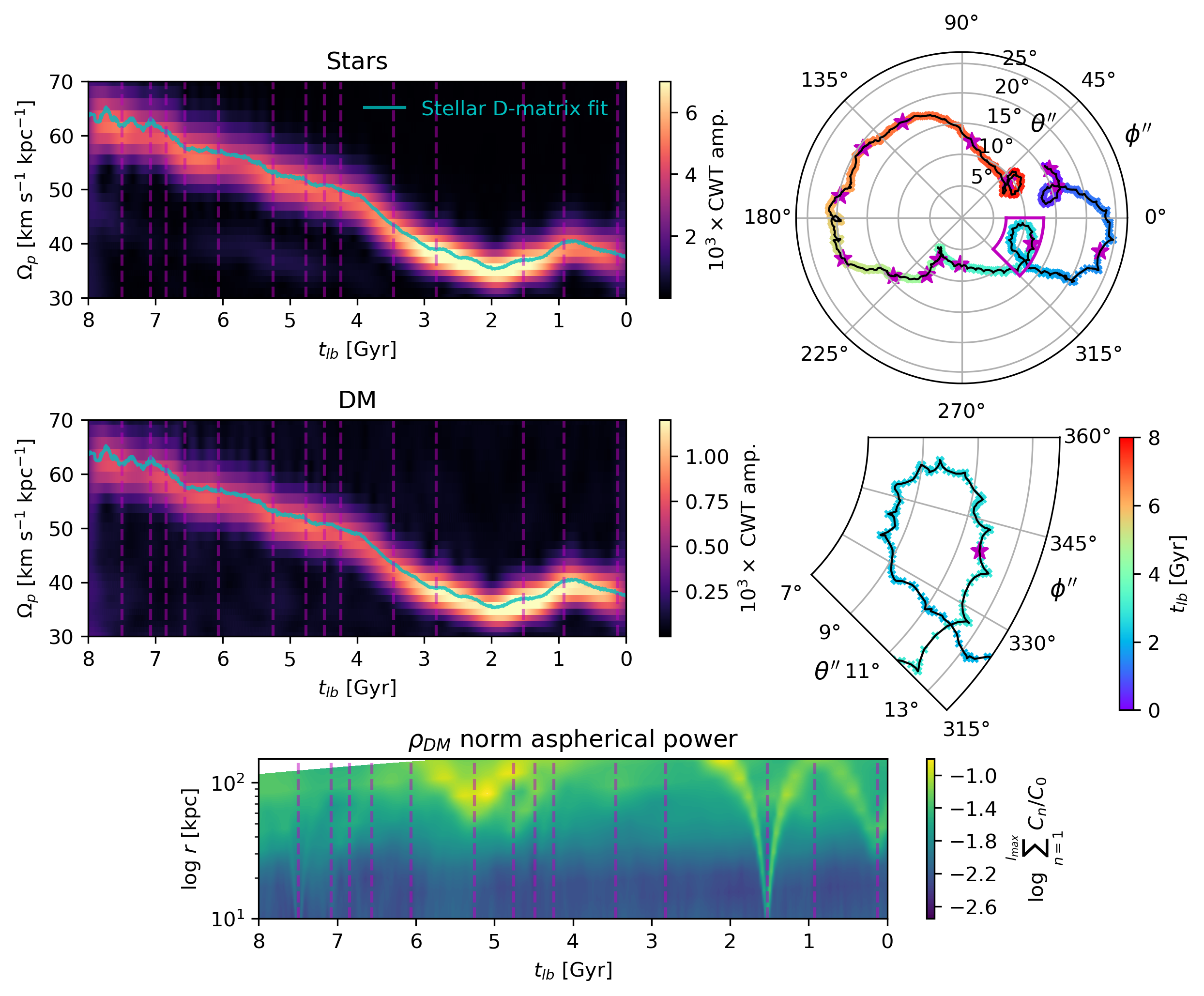}
    \caption{Evolution of the bar pattern speed (left column) and rotation axis orientation (right column) for galaxy 574286 located in Subbox-0 of TNG50 over the last 8 Gyr. The bar pattern speed was measured using both the Wigner $\mathcal{D}$-matrix method using the star particles (cyan line) and by a Continuous Wavelet Transform using the stellar (top) and DM (bottom) particles. The CWT slices shown are taken at $\sim 2.1$ comoving kpc, and show the average CWT amplitude over all 5 quadrupolar BFE terms. The pattern speed of the DM and stellar bars evolve synchronously throughout the last 8 Gyr, and slow down for most of the evolution until they are spun up by a merger beginning around $\sim 2$ Gyr ago. This merger is evident in the power in aspherical terms ($l>0$) of the density BFE in DM (bottom panel), along with several others (approximate pericenter passages marked by the magenta dashed lines). The rotation axis of the bar was determined from the stellar particles using the Wigner $\mathcal{D}$-matrix method and shows both steady precession and nutation throughout the time course, roughly evolving along an angular circle of diameter $\sim 45^\circ$. The precession frequency varies over the time course, at times rising suddenly (see e.g., the blow-up of a loop structure on the bottom right) which may correspond to merger events (see bottom panel) where the bar could be subjected to sudden torquing from a massive body outside of the bar rotation plane. We mark the times of satellite pericenter passages in this plot as magenta stars. Nutations are visible as the higher frequency arcs which occur repeatedly throughout the loop structure and continuously during the time course. The rotation axis precession is shown in an arbitrary inertial reference frame in which $\phi''$ and $\theta''$ refer to the azimuthal and polar angles respectively.}
    \label{fig:pattern_speed_and_rotation_axis_evolution}
\end{figure*}

To examine the evolution of the bar pattern speed over the $\sim 8$ Gyr duration, we use both a CWT analysis on the BFE coefficients and our Wigner $\mathcal{D}$-matrix method, introduced in section \ref{methods:WignerDmat}. We perform our CWT simultaneously over time and radius for each of the quadrupolar BFE terms for both stellar and DM BFE coefficients, and average over the quadrupolar $m$ moments to account for the arbitrary orientation of our reference frame. CWT results for the stellar and DM BFE coefficients are shown in  the top and bottom of the left column of figure \ref{fig:pattern_speed_and_rotation_axis_evolution}, respectively. The CWT amplitude shows remarkable agreement with the pattern speed fit performed using our Wigner $\mathcal{D}$-matrix method, which falls nearly exactly along the peak amplitude of the CWT at each point in time. This is not a trivial result; the CWT measures frequencies only in the coefficients of the quadrupolar spherical harmonic functions, which have differing orientations and are not necessarily aligned to the plane of the bar's rotation. In contrast, the Wigner $\mathcal{D}$-matrix fit takes advantage of all 5 coefficients to determine the pattern speed within the plane of rotation at each time.

The DM and stellar bars share a common pattern speed throughout the duration of the 8 Gyr time course, as shown by our CWT analysis. While the amplitude of the CWT is a factor $\sim 6$ higher on average for the stellar coefficients, the DM and stellar CWT show a nearly identical pattern speed evolution, and the pattern speed fit using our $\mathcal{D}$-matrix method with the stellar particles precisely matches the central frequency for both the DM and stellar bars.  

Throughout much of its evolution, we find that the DM and stellar bars show a decreasing pattern speed. The bar is impacted by several mergers, with 5 larger mergers whose pericenter passages are around 7.5, 5.2, 4.8, 1.5, and 0.1 Gyr ago (see the bottom panel of figure \ref{fig:pattern_speed_and_rotation_axis_evolution}). The merger 1.5 Gyr ago appears to deliver additional angular momentum to the bars, causing the pattern speed to increase from $\sim 35$ to $\sim 40$ km s$^{-1}$ kpc$^{-1}$ over roughly 1.2 Gyr. The earlier mergers do not appear to cause a significant changes to the pattern speed, though oscillations in the Wigner $\mathcal{D}$-matrix fit are visible between 7-8 Gyr ago.

We find that the bar rotation axis is not static in time, rather it undergoes both precession and nutation over the full 8 Gyr time course. We show this behavior in the polar plot of figure \ref{fig:pattern_speed_and_rotation_axis_evolution}. Neither the precession or nutation frequencies are constant in time, as demonstrated by the change in curvature of the track describing axis orientation evolution. At a few different points in the time course, the precession frequency becomes quite large, forming closed ``loops'' in the track (see the cutout at the bottom of the  right-hand column). The disk short axis outside the bar region undergoes a similar amount of precession to the bar rotation axis over the time course. However, the disk axis and bar rotation axis remain misaligned over the full 8 Gyr by 5-10 degrees. Therefore, some but not all of the rotation axis evolution can be accounted for by global tumbling of the disk.

Determining the exact cause of the rotation axis evolution is beyond the scope of this work. Assuming that the bar's rotation is well approximated by a solid body, there are likely contributions from both torque-free precession and torques which are misaligned with the bar's angular momentum axis. This second scenario may especially be relevant during mergers, when perturbations from massive bodies outside of the disk plane are likely, however the bar may be torqued by the disk itself when the bar's rotational plane is misaligned with the disk plane. Notably, many or possibly all of the ``loop'' structures seen in the rotation axis track coincide with satellite mergers (see fig \ref{fig:pattern_speed_and_rotation_axis_evolution}). 

\subsubsection{Bar shape evolution}

\begin{figure*}
    \centering
    \includegraphics[width=\textwidth]{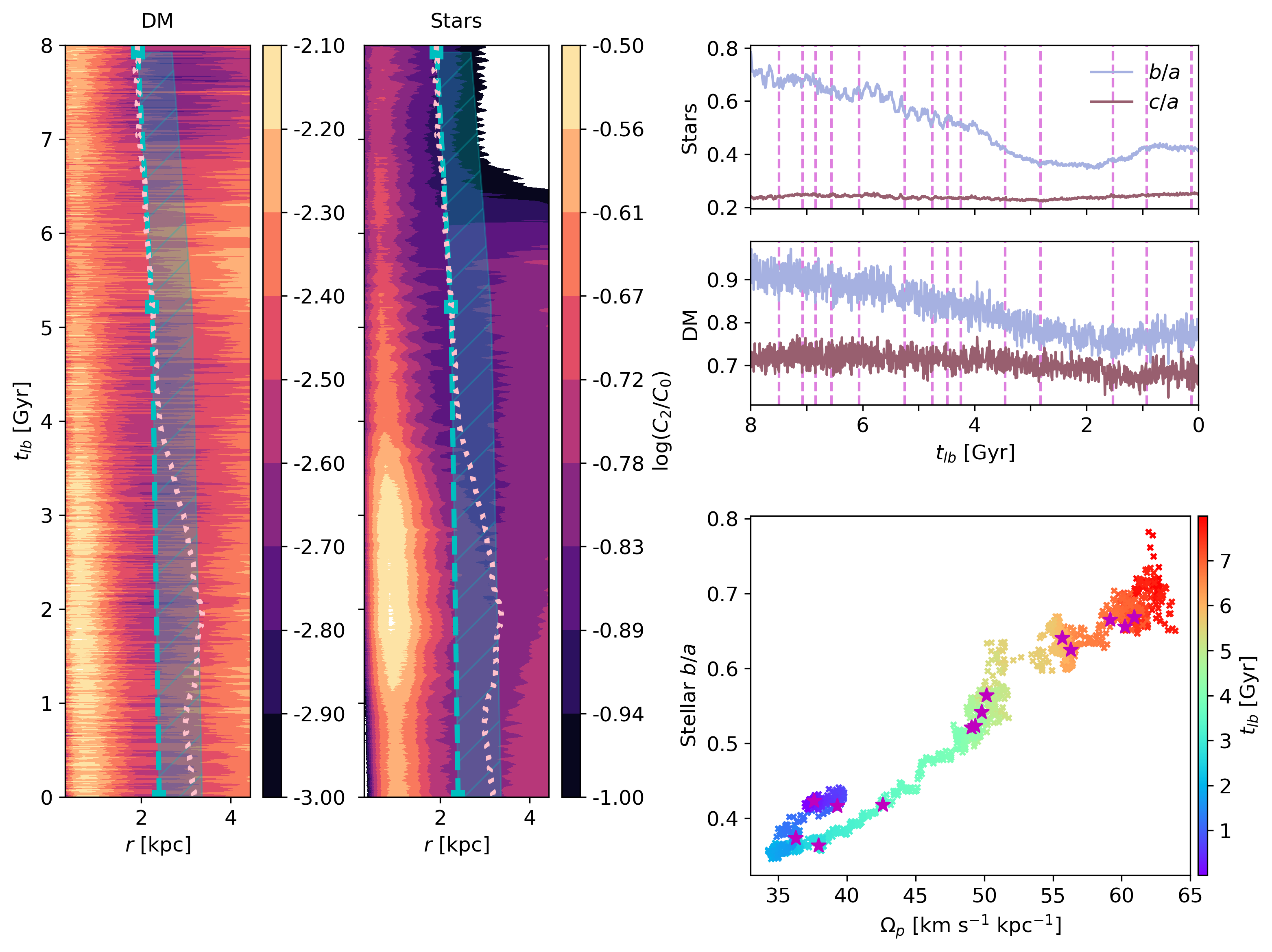}
    \caption{Evolution of bar shapes. The normalized spherical harmonic quadrupole power $C_2/C_0$ (left columns) demonstrate the presence of the bar in both DM and stars by the relatively large amplitude at small radii, which falls to some minimum value around $\sim 3$ kpc before rising again due to the disk/halo component. $C_2/C_0$ does not share a common minimum with the Fourier $A_2(r)$ (bar length as measured by \citetalias{rosas-guevara_evolution_2022}, cyan squares at $t_{lb} = 0,5.2,$ and $7.9$ Gyr) The corotation radius of the bars is overplotted (pink dotted line), demonstrating that the stellar bar is always a fast bar ($r_{bar}\leq r_{corot} \leq 1.4\cdot r_{bar}$, shaded cyan region shows $1-1.4 \cdot r_{bar}$). We track the evolution of the triaxial intermediate:major ($b/a$) and minor:major ($c/a$) axis ratios of the DM and stellar mass distributions measured at 1 kpc (top right). While the DM axis ratios are higher than the those of the stellar bar at all times, in both cases the $b/a$ ratio decreases for much of the evolution while the $c/a$ ratio remains relatively constant. The evolution of the $b/a$ axis ratio is coupled to the bar pattern speed evolution, as shown by the panel in the bottom right. In the $b/a-\Omega_p$ plane, the stellar bar is seen to evolve steadily along a $\sim$linear track, with its pattern speed decreasing as the bars become more prolate in the plane of rotation. The impact of mergers are visible is this plot as vertical strips around $t_{lb}\sim 8$, $5.5$, and $1.5$ Gyr where there are sudden changes in the bar shape without much change in pattern speed. As in figure \ref{fig:pattern_speed_and_rotation_axis_evolution}, we mark the times of satellite pericenter passages with vertical dashed magenta lines and magenta stars.}

    \label{fig:bar_size_evolution}
\end{figure*}

It is not possible for us to follow conventional bar length measurements based on $A_2(r)$ to measure the length of a DM bar. \citetalias{rosas-guevara_evolution_2022} calibrate their definitions for the existence and radius of a bar based on the $A_2(r)$ values for stellar bars, but because DM bars are far more spherical and their $A_2(r)$ are far lower than those for stellar bars, these criteria are not appropriate for DM bars. The analog to the $A_2$ Fourier moment is the normalized quadrupole power, $C_2/C_0$, which is defined as 
\begin{equation}
    C_2/C_0 = \frac{1}{2l+1}\sum_{m=-l}^l\frac{(\rho_l^m)^2}{(\rho_0^0)^2}\bigg|_{l=2},
\end{equation}
is correlated with $A_2$ but still does not allow a simple measurement of bar length. We show its behavior as a function of radius and time in the left two panels of figure \ref{fig:bar_size_evolution}. Notably, $C_2/C_0$ shows a large amplitude out to roughly $\sim 2$ kpc due to the bars in both the stellar and DM populations, in rough analogy to $A_2$ (see the left side of figure \ref{fig:bar_size_evolution}. The bar lengths measured by \citetalias{rosas-guevara_evolution_2022} (over-plotted in cyan) at $t_{lb} = 0$, $5.2$, and $7.9$ Gyr correspond roughly to where $C_2/C_0$ approaches its minimum just beyond the bar region for both DM and stars, whereas they are defined to occur where $A_2(r)$ is at a minimum (if the minimum exists). Because of the lower $A_2(r)$ amplitudes in DM bars and the apparent discrepancy between $C_2/C_0$ and $A_2(r)$, we do not attempt to define a metric to determine the length of a DM bar in this work.

Instead of attempting to measure the time evolution of the DM and stellar bar lengths, we track the evolution of the bar shape in terms of its intermediate:major $b/a$ and minor:major $c/a$ axial ratios, determined using the iterative shape tensor method with an enclosed volume of radius 1 kpc, corresponding roughly to where the bar $A_2(r)$ amplitude is at its maximum. The time series of the three principal axis ratios for both the stellar and DM bars are shown in the upper right of figure \ref{fig:bar_size_evolution}. We note that this is not itself a measure of the axis ratios of the bars, but rather a measure of the axis ratios of the mass distributions within roughly 1 kpc. For this reason, at early times the $b/a$ axis ratios may be higher if this volume extends beyond the bar into the disk/halo region, where the mass distribution is more axisymmetric. 

We find that the axis ratios for both the DM and stellar distributions become more prolate ($b/a$ is lowered) over much of their evolution, while their $c/a$ ratio is largely unchanged. Impacts of mergers are visible in the $b/a$ ratios. In the stellar distribution, oscillations in the $b/a$ ratio are visible between $6-4.5$ Gyr ago, which may indicate a ``ringing'' response to a merger which occurs during this time (We show the influence of this merger in the bottom panel of \ref{fig:pattern_speed_and_rotation_axis_evolution}). Beginning around $t_{lb} = 3$ Gyr, the steady decline in the $b/a$ axis ratios is halted, and $b/a$ begins to increase around $t_{lb} = 2$ Gyr in both DM and stellar distributions. 

The evolution of the $b/a$ ratio is evidently coupled to the pattern speed evolution, illustrated in the bottom right panel of figure \ref{fig:bar_size_evolution}. We find that the bar evolves on a nearly linear track in the $b/a-\Omega_p$ plane, with lower pattern speeds corresponding to a more prolate bar shape in the plane of rotation. This picture is consistent with the classical picture of the bar growing in strength by transporting angular momentum into the outer halo \citep{debattista_constraints_2000,athanassoula_bar-halo_2003}. This evolution is evidently disrupted by the late merger between $1-3$ Gyr ago, which raises the $b/a$ ratio in both DM and stellar bars, and increases the pattern speed. It is likely that this merger is responsible for the increase in bar pattern speed, possibly by delivering gas to the disk which can contribute angular momentum to the bar \citep{beane_stellar_2023}. 

\section{Discussion and Conclusions}

We have shown using a catalogue of disk galaxies in TNG50 that DM ``halo'' bars are common in the presence of stellar bars, and that these DM bars are aligned with their stellar counterparts. DM bars have been previously observed to form in isolated $N-$body simulations of disk galaxies with live halos. \cite{ berentzen_growing_2006} demonstrate the formation of DM bars in cosmologically motivated halos and \cite{marostica_response_2024} form DM bars in the presence of hydrodynamically modelled gas, but to the best of our knowledge DM bars have not been previously noted in cosmological simulations where mergers, substructure, and gas physics are each modelled. The presence of these bars is readily identified in BFE models of the DM potential as semi-sinusoidal oscillations of the quadrupole terms in the inner $\sim 3$ kpc. Using these BFE components, we have shown that it is possible to measure bar pattern speed and the orientation of its rotation axis, provided a snapshot spacing of order $\sim 10$ Myr. The power in the quadrupolar BFE coefficients may enable a measurement of bar lengths, in analogy to the $m=2$ Fourier amplitude. We leave this determination for later work. 

In one sample galaxy taken from Subbox-0, the shape of the stellar mass distribution in the inner $\sim 1$ kpc appears coupled to the pattern speed evolution, with the $b/a$ axis ratio shrinking as the pattern speed declines. This picture is consistent with a large body of simulations and theoretical work, showing that bars can become longer and slower due to dynamical friction with a trailing DM wake \citep[e.g.][]{tremaine_dynamical_1984,debattista_constraints_2000,athanassoula_bars_2005}. Indeed, we observe evidence for a trailing wake structure extending beyond the DM bar in the quadrupolar BFE coefficients of the DM potential (see figure \ref{fig:l2_coeffs_and_contours}). Both the DM and stellar bars rotate very fast ($\sim 60$ km s$^{-1}$ kpc$^{-1}$) at early times ($t_{lb}=8$ Gyr), and show a steady decline to $\sim 35$ km s$^{-1}$ kpc$^{-1}$ around 2 Gyr ago. At this stage, the host halo is impacted by a merger, and both the DM and stellar bars coherently spin up some $\sim 5$ km s$^{-1}$ kpc$^{-1}$, to roughly $40$ km s$^{-1}$ kpc$^{-1}$. The merging satellite may have delivered gas into the bar region, which could provide a positive torque on the bars and possibly cause the observed spin up \citep{sellwood_barhalo_2006, beane_stellar_2023}. The pattern speed does not reach a steady-state after being spun up, but instead begins to slow again roughly 800 Myr ago at a rate nearly identical to the pre-merger slowdown rate. At no point in the time course does the pattern speed dip below 34 km s$^{-1}$ kpc$^{-1}$. 

Because DM bars appear to form in the presence of stellar bars when a live $N$-body halo is modelled in both idealized and cosmological simulations, we cannot gauge the impact of the DM bar on angular momentum transport in this work. However, we find no evidence in TNG50 that DM bars lag behind their stellar counterparts. This suggests that in $\Lambda$CDM DM bars are unable to exert a negative torque on the stellar bar to contribute to pattern speed slowdown as has been previously suggested \cite{athanassoula_bar_2007}. Indeed, \cite{petersen_dark_2016} find that prohibiting the formation of the DM bar in their simulations enhances the rate of bar pattern speed slow-down. Negative torques on the bar are most likely caused by near-resonant material in the trailing DM wake. Some of this material can become trapped in bar-like orbits, prohibiting it from negatively torquing the bar.


We trace the detailed evolution of the rotation axis orientation of the stellar bar in our sample galaxy and observe both precession and nutation which occur continuously and evolve over the full 8 Gyr time course, with remarkable similarity to rigid-body rotation dynamics. This precession is long-lived and does not decay, nor does the rotation axis tend to a fixed position. We see several distinct points at which the frequency of precession visibly rises, in some cases forming closed ``loop'' structures in the axis orientation track. We suggest that these loops likely correspond to times during which the inner halo is impacted by mergers, when the bars may experience significant off-axis torques. We save further investigation of these torques, rotation axis evolution, and the DM wake structure for future work.




\begin{acknowledgments}
We thank the IllustrisTNG collaboration for providing access to the TNG50 simulation data and their virtual jupyterlab workspace. We additionally would like to thank Behzad Tahmasebzadeh, Eugene Vasiliev, and Leandro Beraldo e Silva, for numerous suggestions and illuminating discussions which helped make this work possible. MV \& NA gratefully acknowledge financial support from NASA-ATP award  80NSSC20K0509. MV also acknowledges support from the National Science Foundation grant AST-2009122.
\end{acknowledgments}

\software{\texttt{astropy} \citep{the_astropy_collaboration_astropy_2018},  
\texttt{AGAMA} \citep{vasiliev_agama_2019}
\texttt{numpy} \citep{harris_array_2020}, 
\texttt{scipy} \citep{virtanen_scipy_2020},
\texttt{quaternionic} \url{https://quaternionic.readthedocs.io/en/latest/},
\texttt{spherical} \url{https://pypi.org/project/spherical/},
\texttt{PyWavelets} \citep{lee_pywavelets_2019}.        
          }
          
\bibliography{references}{}
\bibliographystyle{aasjournal}



\end{document}